\documentclass{article}

\usepackage[utf8]{inputenc}
\usepackage{epsfig}
\usepackage{amsfonts}
\title{Vector fields on C*-algebras, semigroups of endomorphisms and gauge groups}

\usepackage{amsmath}

\author{ Innocenti V. Maresin }
\newcommand{\A}{{\mathcal A}}
\newcommand{\Aherm}{{\A_{herm}}}
\newcommand{\I}{\mathbf{1}}
\ifx\Id\undefined\newcommand{\Id}{\mathrm{Id}}\fi

\newcommand{\As}{{\A_V}}

\newcommand{\mul}{\hspace{0.125em}}
\newcommand{\com}[2]{{[{#1},{#2}]}}

\newcommand{\diff}{{\mathrm{d}}} 
\newcommand{\VF}{V}
\newcommand{\conj}[1]{{#1}^*}
 
\newcommand{\D}{{\mathrm J}} 
\newcommand{\spinor}{L}
\newcommand{\spinorconj}{{L'}}
\newcommand{\flux}[1]{\dot{#1}} 

\newcommand{\Hi}{{\mathcal H}}
\newcommand{\Hx}{{\mathcal D}} 
\newcommand{\CS}{S}

\newcommand{\ddt}{{\frac{d}{dt}}}

\DeclareMathOperator{\cpos}{\circ}
\DeclareMathOperator{\diag}{{\mathrm{diag}}}

\newcommand{\E}[1]{{E^{#1}}}
\newcommand{\EDS}[2]{{\E{\displaystyle#1}_{_{\displaystyle\ #2}}}} % _{_{ }} is a rendering kludge

\newcommand{\Gge}{{\mathcal G}}
\newcommand{\gge}{{\mathfrak g}}
\newcommand{\ggeR}{{\gge^{\mathbb R}}}
\newcommand{\ggeC}{{\gge^{\mathbb C}}}

\ifx\Ad\undefined\DeclareMathOperator{\Ad}{Ad}\fi
\ifx\ad\undefined\DeclareMathOperator{\ad}{ad}\fi
\newcommand{\AdDS}[1]{{\Ad_{\displaystyle\,#1}}}
\newcommand{\adDS}[1]{{\ad_{\displaystyle\,#1}}}

\ifx\Epsilon\undefined\newcommand{\Epsilon}{{\mathcal E}}\fi

\begin{document}
\maketitle
\begin{abstract}
What is a vector field on a C*-algebra is defined. Its relation to semigroups of *-endomorphisms was researched.
Some results given about those vector fields and semigroups.
There are also various constructions of semigroups including one parametrized by the cone of the future in (3+1)-dimensional Minkowski space.
Physical interpretations will be presented in a separate paper.

% Sections 3 and 4 are not good yet, as of December~3.
\end{abstract}

This work is supported by Russian Foundation for Basic Research,\\grant 10-01-00178.

\subsection{Introduction}
The formalism of C*-algebras is known to be used for quantization and quantum observables for a long time.
Several ideas of differential geometry, such as bundles, differential forms and integrals, admit their translations to algebraic language.
This paper presents one concept which was apparently missing: vector fields.
In Section~1 we define real vector fields on a C*-algebra, a continuous semigroup of *-endomorphisms of the same algebra, the relation between them, and basic examples.
In Section~2 we present some less trivial constructs, such as the commutator of vector fields and a semigroup parametrized by several real variables.
Section~3 gives the construction of vector fields via representations of the algebra.
Section~4 gives some implications to semigroups, if there exist ones generated by vector fields from Section~3.

\subsection{Basic facts about C*-algebras}
C*-algebra~$\A$ is a complex Banach algebra with an antilinear involution, which reverses the order of multiplication:
$\conj{(a\mul b)} = \conj b\mul\conj a$.
The norm must satisfy:
$$ ||a||^2 = ||\conj a\mul a||$$
$\Aherm$ is a real linear subspace of self-conjugate (Hermitian) elements. 

A *-endomorphism does not have to keep the norm. Though, one can prove that a *-endomorphism cannot increase the norm of an element:
$$ ||E(\conj a\mul a)|| = ||E(a)||^2,\ ||E(\conj a\mul a\mul\conj a\mul a)|| = ||E(a)||^4,
\ ||E((\conj a\mul a)^4)|| = ||E(a)||^8,\ \cdots$$

\section{Vector fields and real-parametrized endomorphism semigroups}
\indent1.1. A real-parametrized semigroup (or group) of *-endomorphisms is a family of *-endomorphisms parametrized by non-negative (or real) numbers, such that:
$E_\zeta\cpos E_\eta = E_{\zeta+\eta}$ and $E_0 = \Id$.
These semigroups will be referred simply as “endomorphism semigroups”.
Possible example is $E_t(q) = \exp(-itg)q\exp(itg), g\in\Aherm$, and it is actually the only possible example in finite dimensions. 
If the algebra is infinite-dimensional, then other semigroups may exist, and for a semigroup not defined by a conjugation with an exponent
we cannot compute the derivative~$\ddt E_t(q)$ for any element~$q$.
For most constructions of a semigroup (it's not a theorem), though, we can find such *-subalgebra dense in~$\A$
that *-endomorphisms of a semigroup preserve it and aforementioned derivative is defined for any~$q$ from the subalgebra.
It is obvious that such semigroup, restricted to this subalgebra, satisfies the following ordinary differential problem:
$${\scriptstyle\forall q\in\As:} \ddt E_t(q) = \VF(E_t(q))\,,{\scriptstyle\ \forall t:} E_t: \As\to\As\mbox{\small{} a bounded linear map,}
\ E_0 = \Id, \hspace{1.6em}\mbox{(!)}$$
where $\As$ is a *-subalgebra dense in~$\A$ and $\VF: \As\to\A$ is a fixed (unbounded) linear operator.
Another way to express this is a differential equation $\flux E = V\cpos E$
where the differentiation is understood in the strong operator topology (not norm topology) on bounded linear operators on~$\As$.
Are there necessary conditions on~$\VF$? First, it must commute with~*, because any *-endomorphism has to preserve this operation.
Second, by differentiation of the condition~(!) we obtain the product rule:
$$ \forall u,v \in\As: \VF(u\mul v) = u\mul \VF(v) + \VF(u)\mul v \hspace{3em}\mbox{(\.{ })}$$
In other words, the operator~$\VF$ must be a differentiation on~$\As\,$.

1.2. Definition:
Let $\As$ be a dense *-subalgebra, and $\VF: \As\to\A$ is a linear operator satisfying the equation~(\.{ }).
Then $\VF$ is referred to as a {\em vector field} on~$\A$.
If more precision is required, the one can say that $\VF$ is a vector field on~$\A$ defined in the subalgebra~$\As\,$.
A vector field which commutes with~* will be referred to as a {\em real} vector field.

1.3. Theorem:
Given a real vector field~$\VF$ defined in~$\As\,$, an $\varepsilon > 0$
and a solution of the equation~(!) for~$t\in[0,\varepsilon)$.
Then:\begin{itemize}
\item The solution is a family of *-endomorphisms of~$\As$ (and, hence, of~$\A$ by continuity).
\item The solution, in given conditions, is unique.
\item The solution extends to any $t\in [0,+\infty)$.
\item The solution is continuous by both $q$ (on all~$\A$) and $t$.
In other words, $\E\VF_t$ is continuous in the (strong) operator topology.
\end{itemize}
Note: The restriction that solutions are looked for in an operator space, not as individual trajectories,
allows to bypass uniqueness problems of ordinary differential equations.

Proof. 
From the property~(\.{ }) of~$\VF$ follows a differential equality for a solution~$E$ of~(!):
$$ \forall u,v \in\As: \ddt E_t(u\mul v) = \ddt (E_t(u)\mul E_t(v))$$
By integration of it by~$t$ from $t=0$ we get that $E_t(u\mul v)$ always equals to $E_t(u)\mul E_t(v)$ in~$\As$,
which in turn extends to all~$\A$ by continuity.
Conservation of~* is proved similarly from the condition that $\VF$ is real.
Suppose two solutions $E$ and $\tilde E$. Then, for any $\kappa\in {\mathbb C}$:
$(1-\kappa)\mul E + \kappa\mul \tilde E$ is also a solution of~(!), each of which we already know to be a family of endomorphisms.
% Bounded linear maps from~$\A$ to itself also form a normed vector space. Hence, 
For given~$t$ the dependence on~$\kappa$ can either define a complex line or a fixed endomorphism (if there is no dependence).
But complex lines of *-endomorphisms are impossible: either because they preserve~*, or because each of them has the norm not greater than~1.
%\footnote{
%	It also follows from the fact that a *-endomorphism must preserve self-conjugateness,
%	but there are no complex lines of self-conjugate (Hermitian) elements.
%	Hence, operators has to be independent of~$\kappa$ on~$\Aherm$,
%	but $\A$ is the complexification of~$\Aherm$ as a vector space.
%}
Hence, for any~$t$:
$E_t = \tilde E_t$.
% Another way to prove uniqueness are reasonings based on the theory of topological vector spaces; see 2.6.

Extension to all $[0,+\infty)$ is made by composition, from the Archimedean property of real numbers and uniqueness of the solution.
Continuity by~$t$ for any~$q\in\A$ can be proved from density of~$\As$ and $|E_t(q)|\le|q|$~inequality, for example,
by direct verification of $\varepsilon$–$\delta$~definition of continuity.
[]

Hence, for any real vector field~$\VF$ there are only two possibilities:
either it defines a semigroup of endomorphisms of~$\A$, or the equation~(!) has no solution % (in endomorphisms of~$\As$)
anywhere for $t>0$. The former case will be referred to as “$\VF$ generates a semigroup”, or “$\VF$ is semigroup-generating”,
and this semigroup will be denoted as~$\E\VF$.
Informally, $\E\VF_t$ is the exponential of~$t\mul\VF\,$.

The only obvious property of the set of semigroup-generating real vector fields defined in~$\As$ is that this set is a cone.

1.4. Proposition:
If a vector field~$\VF$ generates a semigroup, then it commutes with all its endomorphisms:
$$ \forall t\ge 0:\ \VF\cpos\EDS\VF t = \EDS\VF t \cpos\VF\,.$$

Proof:
First of all, sides of this equality belong to the same space of operators from~$\As$ to~$\A$,
because $\E\VF$ preserves $\As\,$.
% Let's check it pointwise, for any~$q\in\As\,$:
From~(!):
$$
   \VF\cpos\EDS\VF t = \frac{d}{d\tau}\EDS\VF\tau|_{\tau=t}
 = \EDS\VF t\cpos\left(\frac{d}{d\tau}\EDS\VF\tau|_{\tau=0}\right)
 = \EDS\VF t\cpos\VF $$
where the middle equality follows from the fact that $\E\VF$ is a real-parametrized semigroup of (linear) endomorphisms.
The derivative is understood as pointwise on elements of the *-subalgebra~$\As$.
Or, more “scientifically”, in the strong operator topology from~$\As$ to~$\A\,$.
[]

1.5. Example: let $M$ be a real ${\mathcal C}^1$-smooth manifold.
We'll denote as ${\mathcal C}_0(M)$ the algebra of continuous complex-valued functions vanishing at infinity
with pointwise operations multiplication and * (as the complex conjugation).
“Vanishing at infinity” means that $|f(\cdot)|\ge\varepsilon$ must be compact for any $\varepsilon>0$~– this property is important to get a C*-algebra on a manifold which is not compact.
Then any vector field~$V_M$ on $M$ translates to the vector field on ${\mathcal C}_0(M)$ defined
in the *-subalgebra ${\mathcal C}_0^1(M)$ of continuously differentiable functions (vanishing at infinity with the derivative, if needed)
by tautological pointwise scalar product $$V_{{\mathcal C}^1}(f)(\xi) := (df)(\xi)\mul V_M(\xi) $$
of the 1-form of derivative of~$f$ and the given vector field on~$M$. 
If the differential equation~$\flux\xi = V(\xi)$ has for~$t>0$ the family of solutions depending continuously on the initial condition~$x(0)$,
namely, $F: M\times[0,+\infty)\to M$ such that
$$\forall\xi_0,t: \frac{\partial}{\partial t} F(\xi_0,t) = V(F(\xi_0,t));\ F(\xi_0,0) = \xi_0\,,$$
then the corresponding endomorphism semigroup of~${\mathcal C}_0(M)$ is expressed as:
$$(\E{V}_t(f))(\xi) = f(F(\xi,t))$$
This admits generalizations beyond ${\mathcal C}^1$-smoothness (say, to functions with cusps in a point where $V$ is not defined)
with appropriate tuning of the *-subalgebra.

% not critical for the first version
%If $M$ is a manifold with boundary, then the *-subalgebra has to be tuned to generate a semigroup
% depending on the direction of~$V_M$ at each point of boundary (either outwards or inwards).
%
%% Namely, for any point $\sigma \in dM$ there must be $(d\varphi)(\sigma)\mul V_M(\sigma) \le 0$, where $\varphi$ is a function which locally defines the boundary: $\varphi = 0$ on $dM$, $\varphi > 0$ on the interior of~$M$, and $d\varphi\neq 0$.
\noindent\parbox[b]{212pt}{
\vspace{1ex}
\hspace{16pt}1.6. Theorem: for the commutative C*-algebra~${\mathcal C}_0({\mathbb R})$ of continuous functions vanishing at infinity
and its *-subalgebra~${\mathcal C}_0^1({\mathbb R})$ of continuously differentiable functions,
the set of semigroup-generating real vector fields defined in~${\mathcal C}_0^1({\mathbb R})$ is not convex.
\\\vspace{1ex}
\hspace{16pt}Proof:
$$V_1 = (\sqrt[3]{x}+1)\frac{d}{dx},
\ V_2 = (\sqrt[3]{x}-1)\frac{d}{dx}
$$
Both vector fields generate semigroups, and even groups.\\
For~$x>-1$ it may be expressed with a ${\mathcal C}^1$-substitution:
$$ y := \int\frac{1}{V_1} = \int\frac{dx}{\sqrt[3]{x}+1} = 3(\ln(\sqrt[3]{x}+1) - \sqrt[3]{x} + \frac{1}{2}x^{\frac23}),
\ y \in{\mathbb R}$$
which gives $V_1 = {d}/{dy}$ and $(\E{V_1}_t(f))(y) = f(y+t)$.
Note that the ${\mathcal C}^1$-smoothness of~$y$ relatively to~$x$ in $x=y=0$ is not obvious in the given expression, but can be verified by differentiation by~$x$.
The similar expression (infinitely smooth this time) exists for the action of~$\E{V_1}$ on $x<-1$.
$V_2$ gives same substitutions save for replacement of~$x$ with~$-x$ and $x>-1$ with $x<1$.
}\hspace{4pt}\parbox[b]{160pt}{
\epsffile{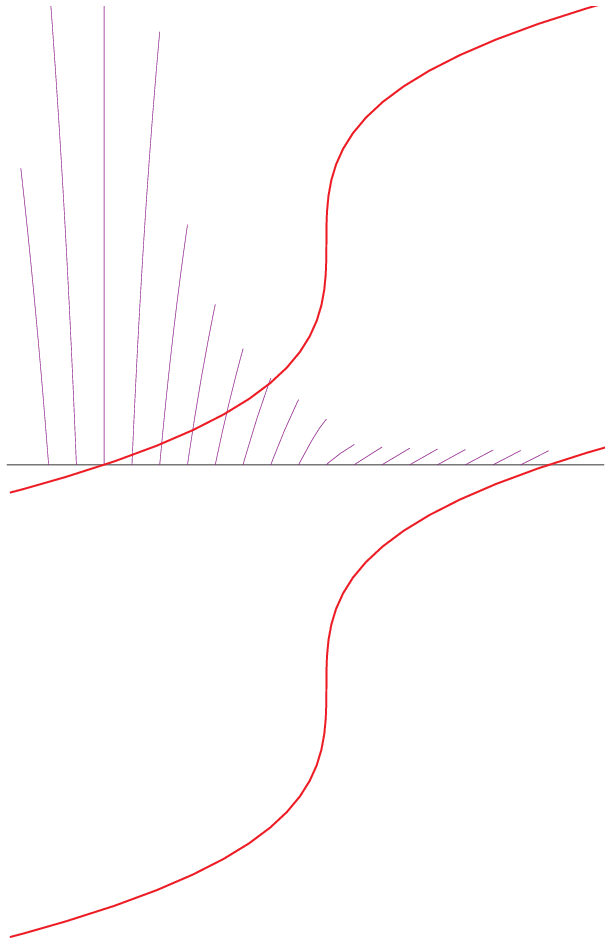}\\
\small
\ Red: graphs $y = V(x)\mul dx$ for $V_1, V_2\,$.\\
\ Purple: integral curves %(approximately)
for~$V_1\,$.% near~$x=-1$.
\\\vspace{1ex}
}\\
\parbox[b]{236pt}{
Indeed, the field $$\frac{V_1 + V_2}{2} = \sqrt[3]{x}\frac{d}{dx}$$is not a semigroup-generating,
since the differential equation $\flux x = \sqrt[3]{x}$ with the initial condition~$x(0) = 0$
% $\flux x = -\sqrt[3]{x}$ can only be solved for~$t \le \frac32 x(0)^{\frac23}$ and this upper $t$~limit vanishes when $x(0)\to 0$.
has many solutions at~$t>0$, for example $x = \pm(\frac23 t)^{\frac32}$,
where the solution with the “+” sign is continuous with solutions for $x(0)>0$
and one with the “$-$” sign is continuous with solutions for $x(0)<0$.
% “−”
This obliges $(\E{(V_1 + V_2)/2}_t(f))(0)$ to take values of~$f$ simultaneously from positive and negative points,
hence such endomorphism is not possible.
[]

Certainly, the same is true for the algebra~${\mathcal C}_0$ of any real interval (of whatever type: closed or open).
}\hspace{8pt}
\parbox[b]{128pt}{
\epsffile{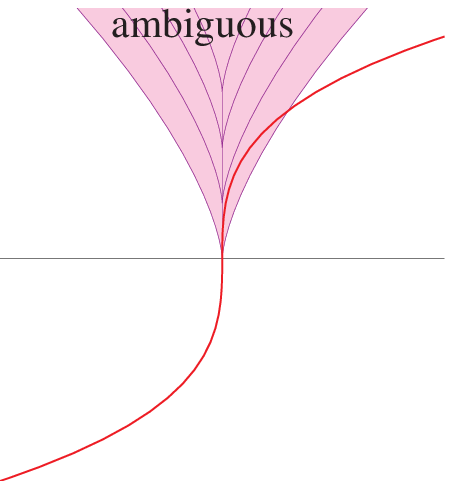}\\
\small
$y = V(x)\mul dx$ for the red curve.
$y = t$ for purple integral curves.
\\
}

Replacement of ${\mathcal C}^1$ with ${\mathcal C}^k$ or ${\mathcal C}^\infty$ does not change much.

The set of semigroup-generating vector fields is not convex even in the simplest case of infinite-dimensional C*-algebra and the most natural choice of a dense *-subalgebra: smooth functions.
Henceforth, one should not expect that this set would be convex in more complicated algebras.
This theorem means that if we are, for some reason, interested in semigroup, then it cannot be constructed easily from generic real vector fields, even if they generate semigroups.

\section{Endomorphism semigroups and gauge groups}
\indent2.1. Note that any element~$g$ of the algebra defines the vector field, defined in all~$\A$,
by adjoint representation as a Lie algebra: $\ad_g(\cdot) := \com{g}{\cdot}$.
Somewhat counter-intuitively, skew-Hermitian ($g+\conj g=0$, i.e. {\em imaginary}) elements
define {\em real} vector fields,\footnote{
	Apparently, the source of well-knows disparateness in real/imaginary factors
	between mathematicians (who follow algebraic intuition)
	and physicists (who might follow differential-geometric intuition).
} and, vice versa, Hermitian elements define “imaginary” (i.e. exchanging Hermitian and skew-Hermitian elements) vector fields,
because the commutator of two Hermitian elements is skew-Hermitian.

Any real vector field on~$\A$ defined by $\ad_g$ (i.e. $g$ must be skew-Hermitian: $g+\conj g=0$)
generates a group of *-endomorphisms.
If the algebra has “1”, then this group can be obtained via exponentiation.
A C*-algebra is a Banach algebra, and we can compute the exponent of an element,
which of skew-Hermitian element will be unitary: $\conj{(\exp(g))}\mul\exp(g) = 1 = \exp(g)\mul\conj{(\exp(g))}\,$.
Unitary elements define (internal) automorphisms of the algebra by conjugation;\footnote{
If there is no “1” in~$\A$, then such automorphisms also exist, %and can be expressed as:
% $$
% \EDS{\ad_g}t(q) = q + \exp_{-1}(t\mul g)\mul q + (q+\exp_{-1}(t\mul g)\mul q)\mul\exp_{-1}(-t\mul g)
% \mbox{ where }\exp_{-1}(x) = \sum_{k=1}^\infty\frac{x^k}{k!},
% $$
although this is not a conjugation in such case.
} see below.

2.2. From here onwards we'll assume that $\A$ is a unital algebra.
Let $\Gge$ be a closed subset of unitary elements, i.e. such $u$ that $\conj u\mul u = 1 = u\mul\conj u$ ,
which forms a group under multiplication.
Then $\Ad_u(\cdot) := u\mul\cdot\mul\conj u$ is its action on~$\A$ by C*-automorphisms.
Denote as~$\gge$ the tangent space of~$\Gge$ in~1. It is easy to check that it is a closed linear subspace,\footnote{
	Over real numbers $\gge$ is a closed subspace of skew-Hermitian elements ($g+\conj g=0$).
	Of course, the complex tangent space is the complexification of the real tangent space.
}
that $\ggeC$ is closed under the commutator operation, but it is not, generally, a *-subalgebra.

Though, exponents of elements of~$\ggeR$ belong to~$\Gge$, because it is topologically closed.

2.3. Definition:
Let $E$ be an endomorphism semigroup, and $\Gge$ be a closed connected group of unitary elements.
If $E_t$ preserves the tangent space~$\gge\ $\footnote{
	It is not important whether $E$ has to preserve real or complex tangent space~$\gge$,
	because any skew-Hermitian element always maps to a skew-Hermitian one by a *-endomorphism.
} for any~$t>0$, then we'll refer to $\Gge$ as a {\em gauge group} for~$E$.

Note that a *-endomorphism (if it is not an isomorphism) usually does not map $\Gge$ to $\Gge$,
even if it preserves its tangent space~$\gge$.

Definition: Let $\VF$ be a real vector field generating an endomorphism semigroup~$\E\VF$,
$\Gge$ be its gauge group, and let the intersection~$\gge\cap\As$ be dense in~$\gge$.
Then we'll refer to~$\Gge$ as a gauge group for the vector field~$\VF$.

In this case, for any $g\in\gge\cap\As$ the derivative $\ddt\E\VF_t(g)|_{t=0}$
(which equals to~$\VF(g)$ by 1.3.) belongs to~$\gge$, which means that
$\forall g\in\gge\cap\As: \VF(g)\in\gge$.
The author currently does not know whether this condition on~$\VF$ is sufficient to have a gauge group~$\Gge$.

2.4. Theorem:
Let $\VF$ be a semigroup-generating real vector field with a gauge group~$\Gge$.
Then:\begin{itemize}
\item The family~$\Ad_u\cpos\E\VF_t$ of *-endomorphisms, where~$u\in\Gge,\ t\ge 0\,$, is a semigroup.
\item For every~$g\in\gge$ the sum of vector fields $\VF+\ad_g$ generates an endomorphism semigroup
which lies within aforementioned two-parameters semigroup.
\end{itemize}

Proof:
Due to associativity of the composition, to prove the first statement it's sufficient to demonstrate that
$$\forall t\ge 0,\ \forall \tilde u\in\Gge: \exists u\in\Gge:
 \AdDS u\cpos\EDS\VF t = \EDS\VF t\cpos\AdDS{\tilde u}$$
Since $\Gge$ is connected, we can check it only in a neighborhood of~1, say, for $||{\tilde u}-1||<1\,$
where natural logarithm is defined by its Taylor series:
$$ \ln(1+x) = x - \frac{x^2}2 + \frac{x^3}3 - \frac{x^4}4 +\cdots$$
After a trivial observation that logarithm of~$\tilde u$ has to belong to~$\gge$
it becomes obvious that $u = \exp(\E\VF_t(\ln{\tilde u}))$ is the solution~–
it directly follows from the fact that $\E\VF_t$ is a *-endomorphism.
We also got that the dependence of~$u$ on both ${\tilde u}$ and $t$ is continuous.

Now we can realize that the semigroup generated by~$\VF+\ad_g$ is expressed as
$$\EDS{\ \VF+\ad_g}t =
 \AdDS{\exp(\int_0^t\EDS\VF{t-\tau}(g)\mul d\tau)_+}\ \cpos\EDS\VF t
\hspace{3em}\mbox{(\char"7E)}$$
where $\exp(\ldots)_+$ is the ordered exponential.\cite{wiki/Ordered_exponential}
% It can be checked against the equation~(!) for~$g\in\gge\cap\As$ and extended for all~$g\in\gge$ by continuity.
For~$0\le s\le t\,$, it satisfies the identity:
$$ \exp(\int_0^t\EDS\VF{t-\tau}(g)\mul d\tau)_+
 = \exp(\int_0^s\EDS\VF{s-\tau}(g)\mul d\tau)_+\mul\exp(\int_0^{t-s}\EDS\VF{t-\tau}(g)\mul d\tau)_+
$$
On the other hand, since $\E\VF_s$ is an endomorphism of the algebra, we have:
$$ \AdDS{\exp(\int_0^{t-s}\EDS\VF{t-\tau}(g)\mul d\tau)_+}
= \AdDS{\exp(\EDS\VF s\,\int_0^{t-s}\EDS\VF{t-s-\tau}(g)\mul d\tau)_+} =$$
$$= \EDS\VF s\cpos\AdDS{\exp(\int_0^{t-s}\EDS\VF{t-s-\tau}(g)\mul d\tau)_+}$$
With it, we immediately see that compositions of *-endomorphisms from~(\char"7E)
behave as they must do it in a semigroup.
So, ordering in the “abnormal” direction, higher semigroup-parameter factors to right and lower to left,
is necessary to put factors with semigroup parameter close to~$t$ adjacent to the $\E\VF_t$~endomorphism.

Initial condition from~(!) is evident.
The only thing we need to complete the proof of~(\char"7E) is to compute its derivative.
Since we already know that the formula~(\char"7E) defines a semigroup, we can do it only at~$t=0\,$:
$$ \ddt\left(\AdDS{\exp(\int_0^t\EDS\VF{t-\tau}(g)\mul d\tau)_+}\ \cpos\EDS\VF t\right)_{t=0} =$$
$$= \adDS{\ddt\exp(\int_0^t\EDS\VF{t-\tau}(g)\mul d\tau)_+|_{t=0}} + \VF = \adDS{g} + \VF\ .$$
[]
\vspace{4ex}

2.5. Commutator of vector fields cannot be defined straightly since we defined the codomain of~$\VF$
to be a broader space than its domain.
Though, if there are two fields $Y: \A_Y\to\A$ and $Z: \A_Z\to\A$
and there exists such dense *-subalgebra~$\A_{YZ}$ that is a subset of both $\A_Y$ and $\A_Z$,
$Y$ maps it into~$\A_Z$ and $Z$ maps it into~$\A_Y$, then the commutator may be defined in this narrower *-subalgebra:
$$\com{Y}{Z} := Y\cpos Z - Z\cpos Y$$

The (\.{ }) condition (see 1.1.) for $\com{Y}{Z}$ is satisfied:
$$\com{Y}{Z}(u\mul v) = Y(u\mul Z(v) + Z(u)\mul v) - Z(u\mul Y(v) + Y(u)\mul v) = $$
$$ = {\scriptstyle
 u\mul Y(Z(v)) + Y(u)\mul Z(v) + Z(u)\mul Y(v) + Y(Z(u))\mul v
- u\mul Z(Y(v)) - Z(u)\mul Y(v) - Y(u)\mul Z(v) - Z(Y(u))\mul v
\ }= $$
$$ = u\mul\com{Y}{Z}(v) + \com{Y}{Z}(u)\mul v\,$$
Since real vector fields commute with~*, it is obvious that
the commutator of two real vector fields is {\em real}; see notes in~2.1.
\pagebreak[4]

2.6. Lemma:
Let $V_1$ be a semigroup-generating real vector field defined in a dense *-subalgebra~$\As$.
% , and having a gauge group~$\Gge$.
Let $V_2$ be a vector field
% defined in~$\As$
 such that
the commutator~$\com{V_1}{V_2}$ is defined in some (narrower) dense *-subalgebra~$\A_{V_1 V_2}$
and is a bounded operator.
Then:
$$ \com{\EDS{V_1}t}{V_2} = \int\limits_0^t \EDS{V_1}\tau\cpos\com{V_1}{V_2}\cpos\EDS{V_1}{t-\tau}\mul d\tau$$

Note that if $\com{V_1}{V_2} = \adDS c$ for certain~$c\in\A$,
then
$$ \int\limits_0^t \EDS{V_1}\tau\cpos\com{V_1}{V_2}\cpos\EDS{V_1}{t-\tau}\mul d\tau
 = \adDS{\int\limits_0^t \EDS{V_1}\tau(c)\mul d\tau}\cpos\EDS{V_1}t
$$
because $\E{V_1}$ consists of endomorphisms of the algebra.

Proof:
First we have to ensure that the operator family~$\com{\E{V_1}_t}{V_2}$ is continuous by~$t$
in the strong operator topology from~$\A_{V_1 V_2}$ to~$\A\,$.
% (which defines a locally convex topological vector space of operators).
It is obvious for the $\E{V_1}_t\cpos{V_2}$~term,
but for~${V_2}\cpos\E{V_1}_t$ we have to check that $\E{V_1}_t$ preserves~$\A_{V_1 V_2}$,
which follows from the boundness of~$\com{V_1}{V_2}$.\footnote{
	More strictly, given $\A_{V_1 V_2}\,$ might have to be extended to some broader *-algebra,
%	possibly even beyond~$\As$ initially given,
	such that both ${V_1}\cpos{V_2}$ and ${V_2}\cpos{V_1}$ are defined
	and the (new) algebra is preserved by~$\E{V_1}$.
}
% By differentiation of the commutator by~$t$ and several algebraic transformations
% (which use Proposition~1.4.)
% we obtain the following ordinary differential equation:
% $$ \ddt\com{\EDS{V_1}t}{V_2} = V_1\cpos\com{\EDS{V_1}t}{V_2} + \com{V_1}{V_2}\cpos\EDS{V_1}t $$
% for which the right side of our statement is a solution.
%% Since left composition with~$V_1$ is a linear continuous operator\footnote{
%%	It is important to get a differential equation with a left composition,
%%	because right composition with an unbounded operator has not such a good property.
%% } in the aforementioned strong operator topology,
%% the solution is unique.
% Note that it is actually the family of (independent) equations on~$\E{V_1}_t\cpos{V_2}(q)$
% for each~$q\in\A_{V_1 V_2}$ where uniqueness follows .
Then, the formula can be checked % for each~$q\in\A_{V_1 V_2}$
by interpolation between ${V_2}\cpos\E{V_1}_t$ and $\E{V_1}_t\cpos{V_2}$
by $\E{V_1}_\tau\cpos{V_2}\cpos\E{V_1}_{t-\tau}$
(say, for $\tau = k\mul t/n,\ k=1\ldots n-1$, like in Riemann integral),
using continuity of~$\E{V_1}_t$ by~$t$.
[]
\vspace{3ex}\strut

2.7. Theorem:
In all conditions of~2.6., let $V_2$ generate its semigroup,
let both semigroups have a gauge group $\Gge$,
and $\exists c\in\ggeR: \com{V_1}{V_2} = \adDS c\,$.\footnote{
	Note that “$\ad$” of any element of~$\A$ is bounded.
%	skew-Hermitian element generates a group of *-endomorphisms
%	because its exponential is unitary and its “$\Ad$” is an internal *-automorphism.
}
Then $ \E{V_1}_t \cpos \E{V_2}_s $ and $ \E{V_2}_s \cpos \E{V_1}_t $ differ only by
left composition with “$\Ad$” of an element of~$\Gge$.
% $$ \com{\EDS{V_1}t}{V_2} = \adDS{C_1(t)}
% \mbox{ where }C_1(t) := ??? $$
% $$ 
%	\AdDS{ false! false! false!
%		\exp(-s\mul\int\limits_0^t \EDS{V_1}\tau(c)\mul d\tau) }\cpos
%	\EDS{V_1}t\cpos\EDS{V_2}s =
%	\EDS{V_2}s\cpos\EDS{V_1}t $$
% blah\mathchar"2D blah
% where $\Gge$-valued function~$C$ is a solution of:
% $$ C(N,0) = 1; $$
% $$ V + N??? = \frac{d}{ds}C(N,V,s) = ??? \mul C(N,V,s)

From this theorem, remembering 2.4.,
immediately follows that the family~$\Ad_u\cpos\E{V_2}_s\cpos\E{V_1}_t$ of *-endomorphisms,
where~$u\in\Gge,\ t\ge 0,\ s\ge 0\,$, is a semigroup.
In a more abstract sense, this means a principal bundle\cite{wiki/Principal_bundle} over~$[0,+\infty)\times[0,+\infty)$
with a semigroup structure corresponding to addition of 2-vectors,
with the representation by~*-endomorphisms and with the left action\footnote{
	The standard definition of a principal bundle requires a right action,
	but for this construct, a left action is better suited.
} of~$\Gge$.
% which is free and transitive on fibers.
It is also possible to prove the statement which generalizes the second statement of the theorem of~2.4.,
but we'll made such generalisation in Section~4 for a more general case.

Proof:
Aforementioned “$\Ad$”~term can be obtained by solving a differential equation (by~$s$) in~$\Gge$,
using Lemma~2.6.\ and the fact that $V_2$ preserves the gauge algebra.
%\vspace{4ex}
[]
% The first equality directly follows from lemma of~2.6., with ... replaced with.

% The second statement follows from~2.4.\ for the vector field~$V_2+\ad_{C_1(t)}$.

%\vspace{19ex}\strut

\pagebreak[3]
\section{Constructions of vector fields}
This section will present possible constructions of vector fields via representations of the algebra.
Although in simple cases these fields generate semigroups, the section does not contain theorems about it.

3.1. Assume that $\A$ has a representation on~$L^2({\mathbb R})$ (with variable~$x\in{\mathbb R}$).
The unbounded skew-Hermitian operator~$\diff = \frac{d}{dx}$ of differentiation by~$x\,$\footnote{
	In quantum-mechanical parlance, $\diff = i\mul\hat{p}/\hbar$.
} is defined on ${\mathcal C}^1$-smooth functions, which are dense in~$L^2$.
Suppose that in some dense *-subalgebra the operator
$\VF(q) = \conj\diff\mul q + q\mul\diff = \com q\diff\,$ is defined.
It means that aforementioned operator on~$L^2({\mathbb R})$ is bounded and belongs to the representation.
For example, such *-subalgebra would be ${\mathcal C}^1_0( {\mathbb R})$ for~$\A = {\mathcal C}_0({\mathbb R})\,$.
From the product rule for~$\VF$ we see that it is a vector field.
If $\A$ is translation-invariant, then $\VF$ generates a group of~*-endomorphisms, called “shifts”.

% If we replace $L^2({\mathbb R})$ with $L^2((-\infty,0])$, then $\VF$ can generate a semigroup,\footnote{
%	For ${\mathcal C}_0((-\infty,0])$ it does, if we let $\As$ be all ${\mathcal C}^1$~functions
%	equal to~0 in~0 with a derivative.
%	Semigroup endomorphisms will be shifts $f(x)\mapsto f(x+t)$ and will fill $(-t,0]$ with zeros.
% }
% but $-\VF$ cannot. (Why?)
%  “$-\diff$” cannot into space.

3.2.
The “translation” vector field $\VF(q) = \conj\diff q + q\diff$ may be expressed in a matrix form:
$$
\D := \left(\begin{matrix} \I \\ \diff \end{matrix}\right);\ \ \ \ {}
\conj\D = \left(\begin{matrix} \I & \conj\diff \end{matrix}\right);\ \ \ \ {}
\VF(q) = \conj\D\mul\left(\begin{matrix} 0 & q \\ q & 0 \end{matrix}\right)\mul\D
$$
Or, more complicated:
$$
\CS := \left(\begin{matrix} 0 & \I \\ \I & 0 \end{matrix}\right);\ \ \ \ {}
\VF(q) = \conj\D\mul\CS\mul\diag(q,q)\mul\D
$$
From skew-Hermitianness of~“$\diff$” we have $\conj\D\mul\CS\mul\D = 0$.

Now let us think that instead of~$\diag(q,q)$ we have another representation~$\pi$ of~$\A$,
in another Hilbert space~$\Hx$, which commutes with certain operator~$\CS$. % such that $\CS^2 = {\mathrm I}\,$.
Suppose that $\D: \Hi\to\Hx$ ({\em transition operator}) is such unbounded operator with dense domain that the operator product
$$ \VF(q) = \conj\D\mul\CS\mul\pi(q)\mul\D $$ for any $q\in\As$
belongs to our (default) representation of~$\A$ in~$\Hi$.

This bears some resemblance to Alain Connes'\cite{Connes} definition of differentiation, by commutator with the operator~$F$,
although we, contrary, try to avoid commutators.

It is easy to check that conditions
$$
\conj\D\mul\CS\mul\D = 0;\hspace{4em}
\forall u,v\in\As: \com u{\conj\D}\mul\CS\mul\com\D v = 0
$$
are sufficient to make $\VF$ a vector field.
Note that $\com u{\conj\D}$ means $u\cpos\conj\D - \conj\D\cpos\pi(u)$, an operator from~$\Hx$ to~$\Hi$,
and similarly $\com\D v: \Hi\to\Hx\,$.
Also, if $\CS$ is self-conjugated, then $\VF$ is real.

3.3. What appears after replacement of one-dimensional “$\diff$” with something like Dirac operator?
Suppose that $\D$ depends linearly on the parameter from~${\mathbb C}^2$.
Say, there are two operators $\D_0, \D_1$ and we define $\D_\phi := \phi^\spinor\mul\D_\spinor\,.$
Then,
$$
{\VF_{\phi\otimes\conj\phi}}(q) = \conj{\D_\phi}\mul\CS\mul\pi(q)\mul\D_\phi
\,;\ \phi\in{\mathbb C}^2
$$
% {\VF_{\phi\otimes\conj\phi}} = {\Conj{\phi^\spinorconj}\mul{\conj\D}_\spinorconj}\mul\CS\mul\pi(q)\mul{\phi^\spinor\mul\D_\spinor}

The use of tensor product symbol is motivated by an observation that the vector field depends on parameters bilinearly:
$$\VF_{\spinor\spinorconj}(q) = {{\conj\D}_\spinorconj}\mul\CS\mul\pi(q)\mul{\D_\spinor}\,,\mbox{ or,}$$
$$
\VF_X(q) = X^{\spinor\spinorconj}\mul{{\conj\D}_\spinorconj}\mul\CS\mul\pi(q)\mul{\D_\spinor}
\mbox{ where }X\mbox{ is a }2\times2\mbox{ matrix.}
$$
We can get a 4-dimensional family of vector fields at the price of only 2-dimensional,
because $\D$ and $\conj\D$ are essentially the same thing.
Though, (for self-conjugated $\CS$) only fields parametrized by Hermitian matrices will be real.
% $\VF_{\phi\otimes\conj\phi}$~fields, and their additive inverses, will be real.
Note that $\phi\otimes\conj\phi$ defines a $2\times2$ Hermitian matrix with rank not greater than~1 and non-negative trace.
We postulated the dependence of~$\D$ on 2 complex numbers, and got a vector field depending on a $2\times2$~matrix,
which gave the structure of (1+3)-dimensional Minkowski space known in Special Relativity.
This construction of the family of vector fields is speculative,
but it leads to interesting situation if we suppose that, in some theory,
we got that namely $\VF_{\phi\otimes\conj\phi} = \conj{\D_\phi}\mul\CS\mul\pi(q)\mul\D_\phi$
are semigroup-generating, but not other fields in this 4-dimensional complex-linear family.

\section{Endomorphism semigroup parametrized by a cone}
\indent4.1. The construction from the previous section gives the~${\mathbb C}^2$ of vector fields of the type~$\phi\otimes\conj\phi$,
which is actually the cone over a Riemann sphere, due to independence of the phase of~$\phi$.
Suppose that all of those fields generate a semigroup.\footnote{
	This is the case, for example, if we defined translations in Minkowski space
	and our representation is defined on functions on the cone of the past,
	or another domain containing the whole cone of the past of each of its points.
	A shift semigroup acts on a domain by filling “overflowed” space with zeros (not explained in this version of the paper).
} Let us also suppose that those fields and semigroups has $\Gge$ as a gauge group and
commutators of those fields lie in the gauge algebra.
%  $$ \com{\D_0}{\D_1} = fck $$  it's a total crap!
% From it, we can see that all those fields commute up to adjoint representation of the gauge algebra,
% just like in conditions of the theorem of~2.7.
% For example, $\com{\VF_{00}}{\VF_{11}} = ???\,$.
% $$ \com{\VF_{\phi\otimes\conj\phi}}{\VF_{\psi\otimes\conj\psi}} = blah\mathchar"2D blah(\phi,\psi)$$
% $$\frux q = {{\conj\D}_\spinorconj}\mul\CS\mul\pi(q)\mul{\D_\spinor}$$

Obviously, this construction corresponds to the geometry of the cone of the past
in the Minkowski space, where 4-coordinates and $2\times2$~matrices correspond as:
$$ X =  
\left(\begin{matrix}
x^0 + x^3	& x^1 - i\mul x^2	\\
x^1 + i\mul x^2	& x^0 - x^3
\end{matrix}\right)
 = x^0\mul{\mathrm I} + \sum\limits_{k=1,2,3}x^k\mul\sigma_k\,,
\mbox{ where }\sigma_k\mbox{ are Pauli matrices.}
$$
Here, the semigroup of translations cannot generated by two 1-parameter semigroups,
so the result from~2.7.\ is not applicable.
BTW, it cannot be generated by {\em any} finite number of 1-parameter semigroups.

How can we generalize 2.7.\ for this geometry?

4.2. The spinor representation possibly gave some insight
about origins of semigroup-generating vector fields parametrized by the light cone
(i.e. by null directions) in the Minkowski space,
but it is useless when we have to prove something about semigroups.
From here onwards we'll forget anything about spinors and think we have $n$ fields~$V_k\,$
and some cone~$\Phi$ in~${\mathbb R}^n$ such that all fields $x^k\mul V_k\,,\ x\in\Phi$ generate a semigroup.
For the example from~4.1.\ it would be
$$ \Phi = \left\{\,x\in{\mathbb R}^4\,|\, x^0\ge 0\,,
\ (x^0)^2 - (x^1)^2 - (x^2)^2 - (x^3)^2 = 0\,\right\} $$
% \ \eta_{\mu\nu}\mul x^\mu\mul x^\nu = 0\,\},
% \ \eta = \left(\begin{matrix}
% 1 & 0 & 0 & 0 \\
% 0 & -1 & 0 & 0 \\
% 0 & 0 & -1 & 0 \\
% 0 & 0 & 0 & -1
% \end{matrix}\right)$$
Let all these semigroups have the same gauge group~$\Gge$.\footnote{
	This version of the paper lacks an explanation
	how gauge groups may be related to the spinor construction from Section~3.
	To be fixed.
}
The condition about commutators takes the form
$$ \com{V_k}{V_l} = \adDS{c_{kl}},\ c_{kl}\in\gge$$
due to bilinearity of the commutator.
Then:
\begin{itemize}
\item There exists a principal bundle~$F\to{\hat\Phi}$ (where $\hat\Phi$ is a convex hull)
with the structure group~$\Gge$
and with a semigroup structure on it which corresponds to vector addition in~$\hat\Phi$.
\item There exists its representation~$\Epsilon$ by *-endomorphisms of~$\A$ which include
all semigroups~$\E{\,x^k\mul V_k},\ x\in\Phi$ such that $\E{\,x^k\mul V_k}_t$ is parametrized
by the point of~$F$ corresponding to the point~$t\mul x^k$ of $\Phi$.
\item The fiber over the 0~point is the group~$\Gge$ itself with its group structure matching
the semigroup structure of the bundle.
% Its right action on the bundle is free and transitive on each fiber.\footnote{
%	This virtually mean that each fiber looks like the action of~$\Gge$ on itself.
% }
% such that it correspond to right composition of *-endomorphisms with~$\Ad_u,\ u\in\Gge$; namely:
% $$ \Epsilon(u(p)) = \AdDS u \cpos \Epsilon(p)\,.$$
So, endomorphisms from the same fiber differ only by left composition with an~$\Ad_u$.
\item All (other) vectors $x\in\hat\Phi$ have their semigroups~$\E{\,x^k\mul V_k}$ included to~$F$ by the same way.
\end{itemize}

Proof:
By applying the theorem of~2.7.\ to two-dimensional subspaces of~${\mathbb R}^n$
generated by pairs of vectors from~$\Phi$ we can construct the bundle over such subspaces.
Where subspaces intersect, consistency is provided by existence and uniqueness
of the corresponding real-parametrized endomorphism semigroup.
Iterating this construction $n-1$~times
(or directly applying 2.7.-like reasonings to $n$-tuples of vectors from~$\Phi$)
we extend it to all~$\hat\Phi$.

The vector fields~$V_k$ (or, in other words, the field linearly parametrized by an $n$-dimentional real space)
actually define the connection on our bundle, where $c_{kl}$ is its 2-form of curvature.
% It is ${\mathcal C}^1$-smooth (why?).
The last (fourth) statement is a manifestation of the fact that such connection % clarify!
% a ${\mathcal C}^1$-smooth connection
can be integrated along the path.
[]

\end{document}